\documentclass[10pt,journal,twocolumn]{IEEEtran} 
\usepackage{filecontents}
\usepackage[noadjust]{cite}
\usepackage{amsfonts}
\usepackage[cmex10]{amsmath}
\ifCLASSOPTIONcompsoc
\usepackage[caption=false,font=normalsize,labelfont=sf,textfont=sf]{subfig}
\else
\usepackage[caption=false,font=footnotesize]{subfig}
\fi
\usepackage{pgfplotstable}
\usepackage{array}
\usepackage{booktabs}

\usepackage{times}
\usepackage{graphicx}
\usepackage{mathrsfs}
\usepackage{amssymb}
\usepackage{stmaryrd}
\usepackage[noadjust]{cite}

\long\def\symbolfootnote[#1]#2{\begingroup%
\def\thefootnote{\fnsymbol{footnote}}\footnote[#1]{#2}\endgroup}
\begin{document}
\title{Optimal Puncturing of Polar Codes \\ With a Fixed Information Set}
\author{
Liping~Li,~\IEEEmembership{~Member,~IEEE}, Wei~Song,~\IEEEmembership{~Student Member,~IEEE}, and Kai~Niu,~\IEEEmembership{~Member,~IEEE}
\thanks{This work was supported in part by National Natural Science Foundation
of China through grant 61501002, in part by Natural Science Project of
Ministry of Education of Anhui through grant KJ2015A102,
in part by Talents Recruitment Program of Anhui University.

Liping Li and Wei Song are with the Key Laboratory of Intelligent Computing and Signal Processing,
Ministry of Education,
Anhui University, China (Email: liping\_li@ahu.edu.cn)}
\thanks{Kai Niu is with Key Laboratory of Universal Wireless Communication, Ministry of Education, Beijing University of Posts and Telecommunications, Beijing 100876, Peoples Republic of China
(niukai@bupt.edu.cn)}
}

\maketitle
\begin{abstract}
For a given polar code construction, the existing literature on puncturing for polar codes focuses in finding the optimal puncturing pattern, and then re-selecting the information set.
This paper devotes itself to find the optimal puncturing pattern
when the information set is fixed.
Puncturing the coded bits corresponding to the worst quality bit channels, called the worst quality puncturing (WQP), is proposed, which is analyzed to minimize the bit channel quality loss at the punctured positions.
Simulation results show that WQP outperforms the best existing puncturing schemes when
the information set is fixed.
\end{abstract}
\section{Introduction}
Polar codes are proposed by Ar$\i$kan in \cite{arikan_it09} and are proven to achieve the capacity
of binary-input, memoryless, output-symmetric (BMS) channels with a low encoding and decoding complexity.
The construction of polar code (selecting the good bit channels from all $N$ bit channels) can be
classified as construction using Monte-Carlo simulations \cite{arikan_it09}, density evolution (DE) \cite{mori_icl09} {\color{black}{\cite{mori_isit09}}},  bit channel approximations \cite{vardy_it13}, {\color{black}{density evolution with a}} Gaussian approximation (GA)
\cite{dai_ia17,p.trifonov_c12}, and polarization
weight (PW) \cite{zhou_vtc18}.

To achieve arbitrary code lengths and code rates, puncturing of polar codes are reported in \cite{eslami_itp11}.
The  quasi-uniform puncturing (QUP) algorithm is proposed in \cite{niu_icc13}.
Re-ordering the bit channels after puncturing with GA is proposed in \cite{zhang_it14}.
The aforementioned puncturing procedures need to re-order the bit channels by GA after puncturing.
In \cite{saber_tc15}, to achieve the maximum throughput, the authors made a  conjecture that
the coded bits with the highest first-error probability should be transmitted first.
As noted in \cite{saber_tc15}, there is no  proof for this procedure.


In practical applications, typically the information bit set is fixed once calculated, for example, in concatenation and interleaving schemes. Therefore, if a system alternates between puncturing and non-puncturing modes, it is desirable to fix the information bit set in order to re-use the existing encoding and decoding structures of polar codes.
The existing puncturing schemes \cite{eslami_itp11,niu_icc13,zhang_it14} are not optimized in this sense.

In this work, puncturing for polar codes is designed to re-use the original information bit set while optimizing the puncturing pattern. We prove that {\color{black}{puncturing the coded bits with indices corresponding to the frozen bit
channels (bit-reversed version)}} is theoretically optimal in terms of the union bound of the block
error probability. We further propose to puncture from the indices corresponding to the worst bit channel quality,
which is called the worst quality puncturing (WQP) in this paper. It is proven that WQP maintains the minimum overall bit channel quality loss at the punctured positions. Note that the work in \cite{zhang_it14} did propose a similar heuristic Algorithm 4.
However, no proof is provided in \cite{zhang_it14} to support this pattern.
Furthermore, in the current paper, the proof does not assume a successive cancellation (SC) decoder in
calculating the likelihood ratios (LRs). Instead, the proof is based on the basic partial order (PO) of polar codes \cite{mori_isit09,li_iwcmc17}, which is independent of the underlying channels or
decoder types.

The remaining paper is organized as follows. In Section \ref{sec.pre1}, the basics of the polar codes are introduced.
The optimal puncture choice is proven from the theoretical perspective in Section \ref{sec_frozen_puncturing}.
The numerical results comparing the WQP algorithm with the exiting puncturing algorithms  are
provided in Section \ref{sec_numerical}. The concluding remarks are at the end.
\section{Preliminary of Polar Codes}\label{sec.pre1}
For a given BMS channel $W$: $\mathcal{X}$ $\longrightarrow$ $\mathcal{Y}$, its input alphabet, output alphabet, and transition probability are $\mathcal{X}=\{0,1\}$, $\mathcal{Y}$, and $W(y|x)$, respectively, where $x\in\mathcal{X}$ and $y\in\mathcal{Y}$.

Let $G_N$ be the generator matrix: $G_N = B_NF^{\otimes n}$, where $N = 2^n$ is the  code length  ($n\geqslant 1$), $B_N$ is the permutation matrix used for the bit-reversal operation,
$F\triangleq[\begin{smallmatrix}1&0\\1&1 \end{smallmatrix}]$,
and $F^{\otimes n}$ denotes the $n$th Kronecker product of $F$.
Let $u_0^{N-1}$ denote a vector containing entries from $u_0$ to $u_{N-1}$.
{\color{black}{A codeword $x_0^{N-1}$} is obtained from $x_0^{N-1} = u_0^{N-1}G_N$.}
A vector channel is defined as:
\begin{align}
W_N(y_0^{N-1}|u_0^{N-1})=W^N(y_0^{N-1}|{\color{black}{x_0^{N-1}}}=u_0^{N-1}G_N)
\end{align}
Splitting $W_N$ to a set of $N$ binary-input channels $W_N^{(i)}$ ($0 \le i \le N-1$), defined as
\begin{align}\label{eq.WN}
W_N^{(i)}(y_0^{N-1},u_0^{i-1}|u_i)=\sum\limits_{u_{i+1}^{N-1}\in \mathcal{X}^{N-i-1}}\frac{1}{2^{N-1}}W_N(y_0^{N-1}|u_0^{N-1})
\end{align}
The channel $W_N^{(i)}$ is called bit channel $i$, meaning that it is the channel that bit $i$ experiences.

The original code  length $N$  of polar codes is limited to the power of two, i.e. $N = 2 ^ n$.
To obtain any code length, puncturing is typically performed.
The code length of the punctured codes are denoted by $M$, containing $K$ information bits. Let $Q$ denote the number of punctured coded bits with $Q=N-M$. The code rate of the punctured codes is $R$ with $R=K/M$.

For the punctured mode, the decoder does not have a priori information of the punctured bits. Equivalently, the transition probability of a punctured channel $H$ is $H (y | 0) = H (y | 1) = 1/2$.
It can be easily shown that the punctured channel capacity is: $I(H) = 0$.


\section{Channel Degradation Process}
\newtheorem{lemma}{Lemma}
The binary expansion of the integer
$i$ ($0 \le i \le N-1$) is: $(i)_b = (b_1, b_2,...,b_n)$ ($b_1$ is the MSB).
Define $\pi \{i\}$
as the bit-reversal permutation operation of $i$:
\begin{equation}
\pi\{i\} = \pi\{(b_1,b_2,...,b_n)\} = (b_n,...,b_2,b_1) =\sum_{k=1}^{n}b_k2^{k-1}
\end{equation}
When the argument of $\pi\{\cdot\}$ is a set, it performs bit reverse to each element
of the set.

The partial order (PO) defined by
Definition 2 of \cite{mori_isit09} is introduced in this section and is used in the sequel to prove the optimum puncturing pattern.
For two bit channels
$i$ and $j$, the $t$th bit of the binary expansions of $i$ and $j$ is {\color{black}{$b_t$}}
and {\color{black}{$b'_t$}}. If for each $t$ ($1 \le t \le n$), {\color{black}{$b_t \ge b'_t$}}, then
bit channel $j$ is stochastically degraded with respect to bit channel $i$, denoted
by $W_N^{(j)} \preceq W_N^{(i)}$ \cite{mori_isit09,li_iwcmc17}.
If $i$ and $j$ satisfy this relationship, we call $j$ is covered by $i$: $j \preceq_c i$.
The degradation from this PO is written as $W_N^{(j)} \preceq_c W_N^{(i)}$ (or $W_N^{(i)} \succeq_c W_N^{(j)}$ )
in this paper
to specifically refer to it, and this PO is written as POc.
Note that 1) a channel is both degraded and upgraded to itself: $W_N^{(j)} \preceq_c W_N^{(j)}$ and $W_N^{(j)} \succeq_c W_N^{(j)}$; 2) the covering and the degradation is transitive. {\color{black}{A simple example shows the POc relationship for
$N=8$. Let $(i)_b = (110)$ and $(j)_b=(100)$. Because for every bit $b_t$ and $b'_t$ ($1 \le t \le 3$) of $(i)_b$ and $(j)_b$,
it holds that $b_t \ge b'_t$. Then the covering relationship exists: $j \preceq_c i$.
With POc, it can be established that $W_8^{(4)} \preceq_c W_8^{(6)}$.}}

\subsection{Basic Degradation Mappings}
The simplest case of POc can be identified for the case of $N=2$: $W_2^{(0)} \preceq_c W_2^{(1)}$, $W_2^{(0)} \preceq_c W_2^{(0)}$, and $W_2^{(1)} \preceq_c W_2^{(1)}$.
With this basic form, a surjective degradation mapping is defined:
\begin{equation}
f:~\mathbb{D} \overset{\succeq_c}{\rightarrow} \mathbb{D}'
\end{equation}
where the domain is $\mathbb{D}=\{\{0\},~\{1\},\{0,1\}\}$ and the codomain $\mathbb{D}' =\{\{0\},\{0,1\}\}$.
The specific degradation mapping of $f$ is:
\begin{eqnarray}\label{eq_N_2_1}
&&\{0\} \text{~or~} \{1\}  \overset{\succeq_c}{\rightarrow} \{0\} \\ \label{eq_N_2_2}
&&\{0,1\} \overset{\succeq_c}{\rightarrow} \{0,1\} \text{~with~} 0 \rightarrow 0 \text{~and~} 1 \rightarrow 1
\end{eqnarray}
For any given element $\mathcal{D} \in \mathbb{D}$, there is an unique element $\mathcal{D}' \in \mathbb{D}'$
obtained from the surjective mapping $f$. Also from (\ref{eq_N_2_1}) and (\ref{eq_N_2_2}), it can be
seen that the cardinality is preserved: $|\mathcal{D}| = |\mathcal{D}'|$. Therefore, the degradation
mapping $f$ is surjective and cardinality preserving. {\color{black}{The mappings defined in (\ref{eq_N_2_1}) and (\ref{eq_N_2_2})
are called the basic degradation mappings.}}

\begin{figure}
{\par\centering
\resizebox*{3.0in}{!}{\includegraphics{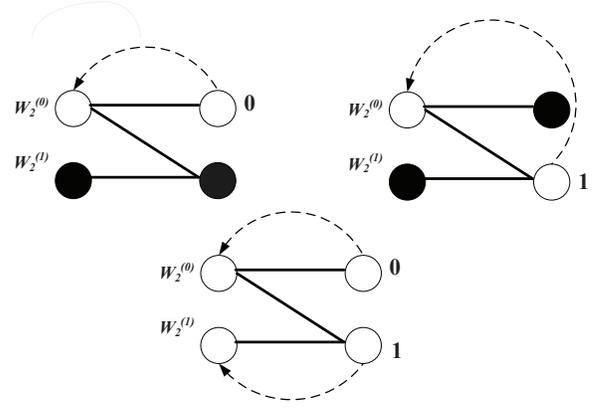}} \par}
\caption{
The basic degradation mappings for $N=2$. The white nodes at the right side are elements in $\mathcal{D}$.
The white nodes at the left side are elements in $\mathcal{D}'$.
}
\label{fig_N_2_degradation}
\end{figure}

{\color{black}{
We provide the following example to
illustrate the mappings defined in (\ref{eq_N_2_1}) and (\ref{eq_N_2_2}). With $N=2$, there are only two
bit channels: bit channel 0 and bit channel 1. Let $\mathcal{D}$ contains any combinations of
these two bit channel indices.
If we want to find out what are the bit channels that are degraded (from POc) to those contained in
$\mathcal{D}$, then (\ref{eq_N_2_1}) and (\ref{eq_N_2_2}) can be employed.
For example, let $\mathcal{D}=\{1\}$. Then
according to (\ref{eq_N_2_1}),  $\mathcal{D}'=\{0\}$. This mapping of course can be immediately verified
since $W_2^{(0)} \preceq_c W_2^{(1)}$. If $\mathcal{D}=\{0\}$, $\mathcal{D}'=\{0\}$ from (\ref{eq_N_2_1}).
This also conforms with the fact that a channel is both statistically degraded and upgraded to itself.
When $\mathcal{D}=\{0,1\}$, the mapping in (\ref{eq_N_2_2}) is applied to obtain $\mathcal{D}'=\{0,1\}$.
In this mapping, (\ref{eq_N_2_2}) also specifies that $0 \rightarrow 0$ and $1 \rightarrow 1$. In
this case, each bit channel in $\mathcal{D}'$ is degraded to one bit channel in $\mathcal{D}$.
The basic degradation mappings can be graphically illustrated in Fig.~\ref{fig_N_2_degradation}.
}}

\begin{figure}
{\par\centering
\resizebox*{3.0in}{!}{\includegraphics{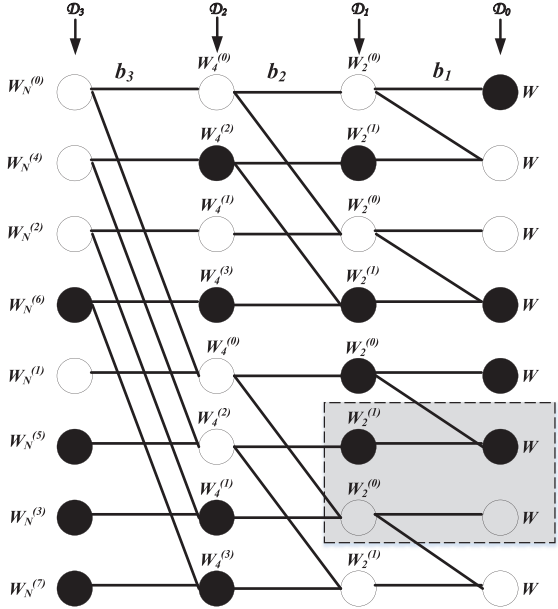}} \par}
\caption{
The bit channel degradation process for $N=8$.  From the right to the left,
there are three levels: level one denoted by $b_1$, level two denoted by
$b_2$, and level three denoted by $b_3$.
}
\label{fig_bit_channel}
\end{figure}

{\color{black}{\subsection{General Degradation Mappings}}}
The mappings defined in the previous section
can be used as the building block to form the degradation mappings for polar codes with a block length
of $N=2^n$. Define a set $\mathcal{D}_0 \subseteq \{0,1,...,N-1\}$. {\color{black}{Like the basic
degradation mappings, a set $\mathcal{D}'$ is desired which contains bit channels which
are degraded (from POc) to bit channels in $\mathcal{D}_0$.}}
This degradation set can be formed
recursively from level $1$ to level $n$ as shown in Fig.~\ref{fig_bit_channel}.
Fig.~\ref{fig_bit_channel} is
a full expansion of the structure of the bit channel transformation of polar codes defined in \cite{arikan_it09}.
The building block in Fig.~\ref{fig_bit_channel} is the one-step transformation of polar codes with $N=2$, illustrated
by the diagram in the gray rectangle: with a bit `0' in the basic block, the output is the upper left channel $W_2^{(0)}$; otherwise the output is the lower left channel $W_2^{(1)}$.
{\color{black}{
When referring to the nodes of the transformation graph, the $N$ nodes in each column are indexed in the
bit-reversed order.
}}

{\color{black}{Since in the general case, the final degradation set $\mathcal{D}'$ is obtained in
$n$ recursive levels, we use $\mathcal{D}_n$, instead of $\mathcal{D}'$, to refer to the final set.
The intermediate set at level $k$ is similarly denoted as $\mathcal{D}_k$.}}
The destination set $\mathcal{D}_n$ from the initial set $\mathcal{D}_0$ can be obtained recursively from the following steps:
\begin{itemize}
\item The entries of the initial set  $\mathcal{D}_0$ are assigned to the nodes of the input of level 1 in the bit-reversed order.
\item Applying the basic degradation mappings in (\ref{eq_N_2_1}) and (\ref{eq_N_2_2}) to level $k$ ($1 \le k \le n$) with
the input set $\mathcal{D}_{k-1}$,  the  set $\mathcal{D}_k$ of level $k$ can be obtained.
\end{itemize}

The process defined in these two steps is called the general degradation process in this paper.
{\color{black}{For example, in Fig.~\ref{fig_bit_channel}, the initial set is $\mathcal{D}_0 = \{2,3,4,7\}$.
The bit-reversed set $\pi\{\mathcal{D}_0\}$ is assigned to the nodes in the right-hand side, denoted by
the white nodes. Following the degradation process, this initial set goes to
set $\mathcal{D}_1$, $\mathcal{D}_2$, and  $\mathcal{D}_3$, all denoted by the white nodes of Fig.~\ref{fig_bit_channel} in
each level. The details of the process are provided below to illustrate the usage of the basic degradation mappings in
the general case.
\begin{itemize}
\item Level 1. The first white node is colored because of $\pi\{4\} = 001$.
The MSB of $(4)_b=(b_1,b_2,b_3)=(100)$ is $b_1=1$. According to equation (\ref{eq_N_2_1}), the reached white node of this basic step is the upper left node, as indicated in Fig.~\ref{fig_bit_channel}. The same mapping is applied to the second white
node. The last two white nodes are the input to the same basic one-step transformation. These two
white nodes are colored because of $\pi\{3\} = 110$
and $\pi\{7\} = 111$.
The MSBs of $(3)_b=(b_1,b_2,b_3)=(011)$  and $(7)_b=(b_1,b_2,b_3)=(111)$ are 0 and 1, respectively.
According to equation (\ref{eq_N_2_2}), both the two nodes at the left-hand side of this connection are white,
as indicated in Fig.~\ref{fig_bit_channel}. With the three basic degradation mapping at level 1, the set $\mathcal{D}_1$
is obtained as: $\mathcal{D}_1 = \{0,2,3,7\}$.
\item Level 2. The basic degradation mappings can be applied to level 2. In this level, the second bit $b_2$ determines the
reached white nodes. The set $\mathcal{D}_2$ can be determined as $\mathcal{D}_2 = \{0,1,2,5\}$.
\item Level 3. In this level, the third bit $b_3$ determines the
reached white nodes. The set $\mathcal{D}_3$ is obtained as $\mathcal{D}_3 = \{0,1,2,4\}$.
\end{itemize}
}}
The following lemma formally states this  general degradation process.
\begin{lemma}\label{lemma_degradation}
Let the set $\mathcal{D}_n$ be obtained from the initial set $\mathcal{D}_0$ following the general degradation process.
For any ${j} \in \mathcal{D}_n$, there is an element $i \in \mathcal{D}_0$ such that $W_N^{(j)} \preceq_c W_N^{(i)}$. The mapping between the sets $\mathcal{D}_n$ and $\mathcal{D}_0$ is unique and $|\mathcal{D}_n| = |\mathcal{D}_0|$.
\end{lemma}
\begin{IEEEproof}
Consider the element $i \in \mathcal{D}_0$ with a binary expansion of $(i)_b = (b_1,b_2,...,b_n)$ ($b_1$ is the MSB).
Let $(i')_b = (b'_1,b'_2,...,b'_n)$ be the index of the node that forms the basic one-step transformation  with
node $i$ {\color{black}{at the input to level 1}}.
{\color{black}{A property of polar code transformation is that at level $k$ ($1 \le k \le n$), the two nodes that form the basic one-step transformation only differ at bit $k$ of their binary expansions \cite{arikan_it09}.}}
Therefore $i$ and $i'$ differ only at bit 1 (the MSB): $b_1 \neq b'_1$.
There are two cases: $i' \in \mathcal{D}_0$ or $i' \notin \mathcal{D}_0$.
{\color{black}{Denote $i_1$ as the index of the reached node of $i$ at level 1, and  the first
bit of its binary expansion is $b^{''}_1$.}}

When $i' \notin \mathcal{D}_0$, then at level $1$, the mapping defined by (\ref{eq_N_2_1}) can be applied:
$i$ is mapped to itself {\color{black}{$i_1 = i$}} ($b_1=0$),
or to $i'$ {\color{black}{with $i_1 = i'$}} ($b_1 = 1$).
Since $b_1 \neq b'_1$,  then $b'_1 = 0$ when $b_1=1$.
Therefore, in this case, {\color{black}{$i_1=i \preceq_c i$ (when $b_1 = 0$) or $i_1=i' \preceq_c i$ (when $b'_1 = 0$ and $b_1=1$), resulting in $b^{''}_1 \preceq_c b_1$}}.

When  $i' \in \mathcal{D}_0$, the mapping in (\ref{eq_N_2_2}) can be applied: $i$ maps to $i$ and $i'$ maps to $i'$.
In this case, $i_1 = i \preceq_c i$, also resulting in $b^{''}_1 \preceq_c b_1$.

{\color{black}{Combining the cases of $i' \notin \mathcal{D}_0$ and $i' \in \mathcal{D}_0$,
 it can be concluded that the reached $i_1$ is covered by $i$ because $b_1$ covers $b^{''}_1$}}.

Recursively applying the mapping in (\ref{eq_N_2_1}) and (\ref{eq_N_2_2})
{\color{black}{ from level 1 to level $k$ ($1 \le k \le n$), and denote $i_k$ as the reached index of $i$ at level $k$.
Following the same reasoning of level 1, the reached index $i_k$ is covered by $i_{k-1}$ ($i_0 = i$) because
the $k$th bit of the binary expansion of $i_k$ is covered by that of $i_{k-1}$. Therefore
}}
 it can be concluded that at level $n$,
the value {\color{black}{$i_n = j$}} that $i$ reaches is covered by $i$ by noting that covering is transitive, which indicates that
$W_N^{(j)} \preceq_c W_N^{(i)}$ from POc.

Since the basic mapping (from $\mathbb{D}$ to $\mathbb{D}'$) in (\ref{eq_N_2_1}) and (\ref{eq_N_2_2}) is unique and cardinality
preserving,
the recursive mapping from level 1 to level $n$ is also unique and $|\mathcal{D}_n| = |\mathcal{D}_0|$.
\end{IEEEproof}

{\color{black}{If we write the nodes in the example of Fig.~\ref{fig_bit_channel} in the order corresponding to the
 basic degradation mappings, the four sets are: $\mathcal{D}_0=\{2,3,4,7\} \implies \mathcal{D}_1=\{2,3,0,7\} \implies \mathcal{D}_2=\{2,1,0,5\} \implies \mathcal{D}_3=\{2,1,0,4\}$. Each element in $\mathcal{D}_k$ is covered by the corresponding
 element in $\mathcal{D}_{k-1}$, as indicated (and can be easily verified) by Lemma \ref{lemma_degradation}.
}}
\section{Optimal Puncturing With a \\Fixed Information Set}\label{sec_frozen_puncturing}

Denote $\mathcal{I}$ as the set
containing the indices for the information bits and the set $\mathcal{F}$ containing the indices for the
frozen bits.


\subsection{Connection of Puncturing with Bit Channel Quality}
When at least one of  the two input channels $W$ is a punctured channel $H$,
the output channel $W_2^{(1)}$ degrades to a punctured channel in the one-step transformation
{\color{black}{\cite{niu_16}}}.
The following lemma states this fact.

\begin{lemma}\label{lemma_puncture_one_step}
With  one of the input channels $W$ being a punctured channel $H$, the bit channel $W_2^{(1)}$
degrades to a punctured channel. If both two input channels $W$ are  punctured channels,
$W_2^{(1)}$ and $W_2^{(2)}$ are punctured channels.
\end{lemma}

The proof of Lemma \ref{lemma_puncture_one_step} follows from the proof of Lemma 1 in \cite{niu_16}.
Lemma \ref{lemma_puncture_one_step} shows that the punctured channel propagation
 of the one-step transformation follows exactly the same unique and cardinality preserving mapping
of the basic degradation defined in (\ref{eq_N_2_1}) and (\ref{eq_N_2_2}). Similar to the degradation
process, a new process, called the puncturing process, can be defined by recursively applying the
one-step puncturing mapping in Lemma \ref{lemma_puncture_one_step}.

With the puncturing process and the degradation process following exactly the same mapping, the following
lemma can be immediately obtained from Lemma \ref{lemma_degradation}.

\begin{lemma}\label{lemma_puncture_effect}
Let the destination set ${\mathcal{Q}_n}$ be obtained from the initial set $\mathcal{Q}_0$ following
the  puncturing process.
Then all bit channels in ${\mathcal{Q}_n}$ are punctured channels.
There is also an unique mapping from $\mathcal{Q}_0$ to
$\mathcal{Q}_n$ with $|\mathcal{Q}_n| = |\mathcal{Q}_0|$.
\end{lemma}

\subsection{Optimal Puncturing Pattern}
The following lemma is one of the main results of this paper.
\begin{lemma}\label{lemma_puncture_frozen}
Fixing a given information set $\mathcal{I}$ and frozen set $\mathcal{F}$, consider a puncturing set $\mathcal{Q}$
with $\mathcal{Q} \subseteq \mathcal{F}$ and another puncturing set $\mathcal{Q}'$ with $\mathcal{Q}'\cap \mathcal{I} \neq \emptyset$.
Then
\begin{itemize}
\item Any selection $\mathcal{Q} \subseteq \mathcal{F}$ is close in the sense that the destination set $\mathcal{Q}_n$
is still a subset of $\mathcal{F}$: $\mathcal{Q}_n \subseteq \mathcal{F}$.
\item When the union bound of the block error probability of the considered system is designed to be smaller than 1/2, then we have $P_B(\mathcal{Q}') \ge P_B(\mathcal{Q})$, where $P_B( \cdot)$ is the union bound of the block error probability
    conditioned on the inside argument.
\end{itemize}
\end{lemma}
\begin{IEEEproof}
For the first part of this lemma, we need to invoke Lemma \ref{lemma_degradation}.
In Lemma \ref{lemma_degradation}, it is shown that  for a bit channel $k$ in $\mathcal{Q} \subseteq \mathcal{F}$,
there is a corresponding bit channel $k'$ in $\mathcal{Q}_{n}$ that is stochastically
degraded to it ($W_N^{(k')} \preceq_c W_N^{(k)}$) because the puncturing
process and the degradation process follows exactly the same mapping.
This indicates that in the construction stage of polar codes, bit channel $k'$ can only be
in $\mathcal{F}$ because $k \in \mathcal{F}$ and $W_N^{(k')} \preceq_c W_N^{(k)}$.
With the unique and cardinality preserving mapping, all bit channels in
$\mathcal{Q}_{n}$ are in the set  $\mathcal{F}$ because  $\mathcal{Q} \subseteq \mathcal{F}$.
This concludes the proof of the first part.

Let $P_b(W_N^{(\{i\})})$ be
the error probability of the $i$th bit channel. Then
the union bound of the block error probability with the set $\mathcal{Q}$ is
\begin{equation}\label{eq_union}
P_B(\mathcal{Q}) = \sum_{i \in \mathcal{I}}P_b(W_N^{(i)})
\end{equation}
From the second part of this lemma, it is assumed that this union bound $P_B(\mathcal{Q})$ is smaller than 1/2.
Now let $i \in \mathcal{Q}' \cap \mathcal{I}$. Let $j \in \mathcal{Q}'_{n}$ be the bit channel that
$i$ reaches at the last level.
From Lemma \ref{lemma_degradation}, it is known that $W_N^{(j)} \preceq_c W_N^{(i)}$.
Since $i \in \mathcal{Q}' \cap \mathcal{I}$, bit channel $j$ is a bit channel that is stochastically degraded to it, which is still possibly a bit channel in $\mathcal{I}$. From  Lemma \ref{lemma_puncture_effect}, it is known that the propagated sets $\mathcal{Q}_{n}$ and
$\mathcal{Q}'_{n}$ contain  punctured channels. In the case that $j \in \mathcal{Q}'_n$ and $j \in \mathcal{I}$,  the puncturing set
$\mathcal{Q}'$ renders an information bit channel $j$ to be a punctured channel, which bears an error
probability $P_b(W_N^{(j)}) = 1/2$. Seen from (\ref{eq_union}), the union bound  $P_B(\mathcal{Q}') > 1/2$.
Therefore,  $P_B(\mathcal{Q}') \ge P_B(\mathcal{Q})$ considering the fact that
$\mathcal{Q}_{n} \subseteq \mathcal{F}$ and  $j \in \mathcal{Q}'_{n}$ is possibly in the set $\mathcal{I}$.
\end{IEEEproof}

Before introducing Lemma \ref{lemma_wqp}, the bit channel quality loss is defined for each punctured
coded bit. Let the index of a punctured coded bit be $\pi\{i\}$ with $i \in \mathcal{Q}$.
And let the unique mapping
of this punctured bit be $j \in \mathcal{Q}_n$.
{\color{black}{ According to Lemma \ref{lemma_puncture_effect}, bit channel $j$ is also a punctured channel.
For a punctured channel, the error probability is therefore 1/2 as can be
seen from Section \ref{sec.pre1}.
Let $P_b(W_N^{(j)})$ be the error probability of bit channel ${j}$ without puncturing.
The bit channel quality loss $\alpha_i$ is the difference of the bit channel error probability
before and after puncturing the coded bit $\pi\{i\}$.
}}
Then the bit channel quality loss of puncturing the coded bit $\pi\{i\}$
is defined as:
\begin{equation}
\alpha_i = 0.5 - P_b(W_N^{({j})})
\end{equation}
{\color{black}{Note that}} the error probability $P_b(W_N^{(j)})$ can be obtained
from any of the construction procedures such as \cite{arikan_it09,vardy_it13,dai_ia17,p.trifonov_c12},
which should be available when selecting the information set $\mathcal{I}$.

\begin{lemma}\label{lemma_wqp}
The optimal puncturing pattern $\mathcal{P}_o$ for a given information set $\mathcal{I}$ and frozen set $\mathcal{F}$
must be the one with $\mathcal{P}_o \subseteq \mathcal{F}$. Furthermore, when $\mathcal{P}_o$
contains the bit channels corresponding to the worst channel quality, the selection $\mathcal{P}_o$
produces the minimum bit channel quality loss at the punctured positions compared to all the other puncturing selections.
\end{lemma}

\begin{IEEEproof}
The first part of this lemma is directly from Lemma \ref{lemma_puncture_frozen}. We only need to prove the second part.
Let $|\mathcal{P}_o| = Q$. Ordering the set $\mathcal{F}$ according to the bit channel quality
in the ascending order: $\mathcal{F}=\{f_1,f_2,...,f_{N-K}\}$ and $P_b(W_N^{(f_1)}) > P_b(W_N^{(f_2)}) > ... > P_b(W_N^{(f_{N-K})})$.
Then $\mathcal{P}_o=\{{f}_1,{f}_2,...,{f}_Q\}$,
which contains  indices of the worst $Q$ bit channels.
Consider another puncturing selection $\mathcal{P}_f$ with $|\mathcal{P}_f|=Q$, which replaces
element ${f}_l$ ($1\le l \le Q$) of $\mathcal{P}_o$ with another element ${f}_g$ ($g > Q$).
Then the relationship between these two bit channels
is: $W_N^{(f_{l})} \preceq_c W_N^{(f_{g})}$.
The elements ${f}_l$ and ${f}_g$ propagate to the final level $n$ to
elements ${f}_{ln}$ and ${f}_{gn}$, respectively.
From Lemma \ref{lemma_degradation}, it is known that
$W_N^{(f_{ln})} \preceq_c W_N^{(f_{l})}$ and $W_N^{(f_{gn})} \preceq_c W_N^{(f_{g})}$.
As bit channel $f_{l}$ is among the worst $Q$ bit channels,
then ${f}_{ln} \in \mathcal{P}_o$.
Then with the optimal puncturing $\mathcal{P}_o$, the overall bit channel quality loss at the punctured
positions is:
\begin{equation}
\alpha_{\mathcal{P}_o} = \sum_{i=1}^{Q}\{0.5-P_b(W_N^{(f_{in})})\}
\end{equation}
With $f_g \in \mathcal{F}$ and $W_N^{(f_{gn})} \preceq_c W_N^{(f_{g})}$, it can be concluded that
$f_{gn}$ is still in the set
$\mathcal{F}$: it can be bit channel $f_g$ itself, or a bit channel stochastically degraded to it.
Therefore, $f_{gn}$ is in the set $\{f_1,f_2,...,f_Q, f_{Q+1},...,f_g\}$.
If $f_{gn} \in \{f_1,f_2,...,f_Q\}$, then $\mathcal{P}_f$ reaches to the same set of $\mathcal{P}_o$, which
produces the same overall bit channel quality loss at the punctured position. If $f_{gn} \in \{f_{Q+1},...,f_{g}\}$,
$P_b(W_N^{(f_{gn})}) < P_b(W_N^{(f_{ln})})$ from the ordering of $\mathcal{F}$.
Then overall  it is true that:
\begin{equation}
\alpha_{\mathcal{P}_f} - \alpha_{\mathcal{P}_o} = P_b(W_N^{(f_{ln})}) - P_b(W_N^{(f_{gn})}) \ge 0
\end{equation}
\end{IEEEproof}

The puncturing pattern based on Lemma \ref{lemma_wqp} is called the worst quality puncturing (WQP) in this paper
to differentiate with the existing QUP procedure.

\subsection{Comparison with QUP}
To begin with  the comparison, let us first review the implementation of the QUP scheme.
It works as the following when fixing the information set:
\begin{itemize}
\item For a given number $Q$ of coded symbols to be punctured, set $\mathcal{P}'=\{0,1,...,Q-1\}$;
\item Perform the bit reversal operation to the set $\mathcal{P}'$ to obtain the puncturing
set of the coded symbols $\mathcal{P} = \pi\{\mathcal{P}'\}$.
\end{itemize}
According to Lemma \ref{lemma_wqp}, the WQP works in the following steps:
\begin{itemize}
\item Arrange the frozen set $\mathcal{F}$  in the ascending order of the bit channel quality (as in the proof of Lemma \ref{lemma_wqp});
\item Select the first $Q$ elements of the ordered set $\mathcal{F}$ as the puncturing set $\mathcal{P}_o$;
\item Perform the bit reversal operation to the set $\mathcal{P}_o$ to obtain the puncturing
set of the coded symbols $\mathcal{P} = \pi\{\mathcal{P}_o\}$.
\end{itemize}

\begin{figure}
\centering
\subfloat[QUP]{\includegraphics[width=2.5in]{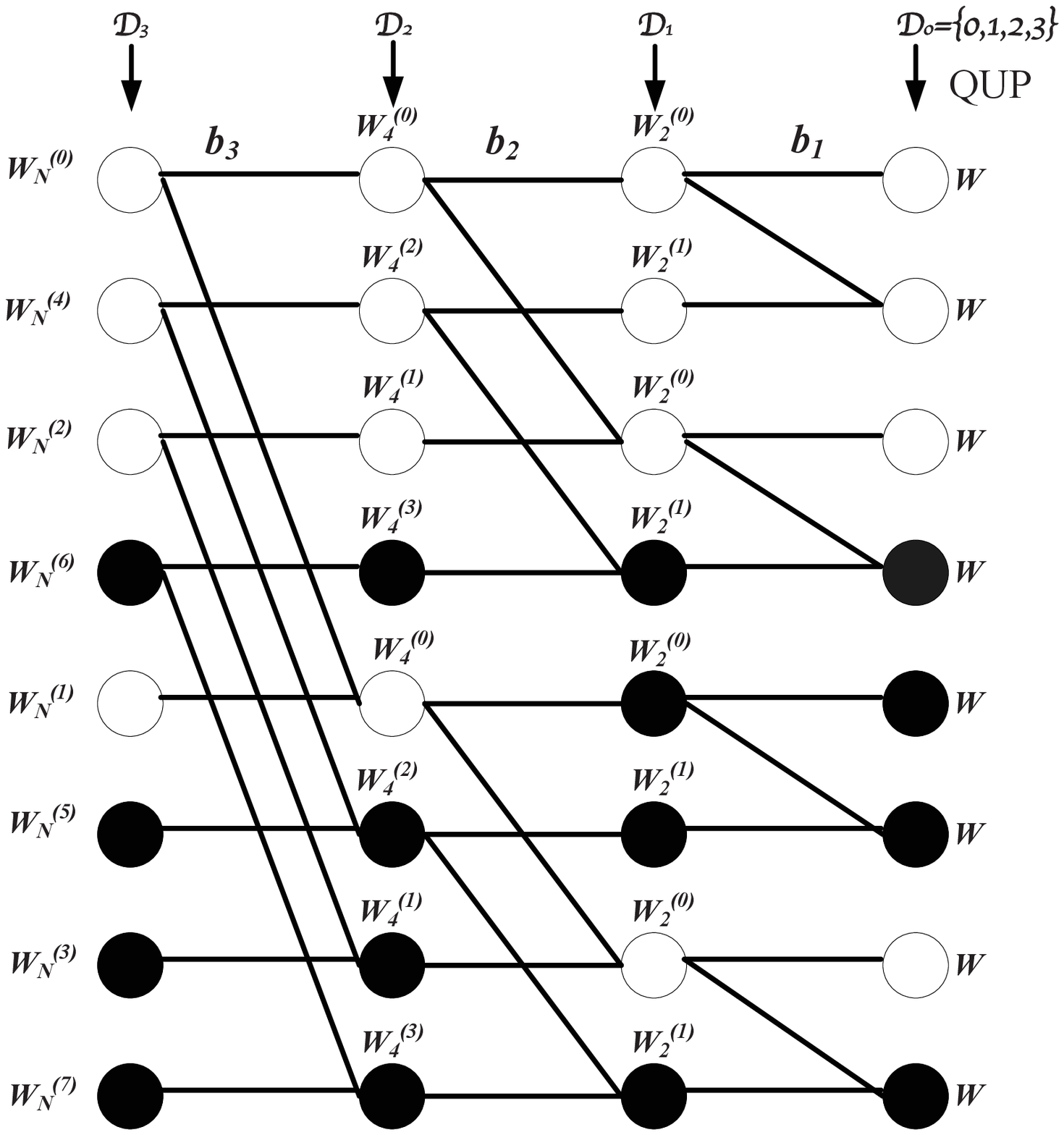}}
\label{fig_qup}
\subfloat[WQP]{\includegraphics[width=2.5in]{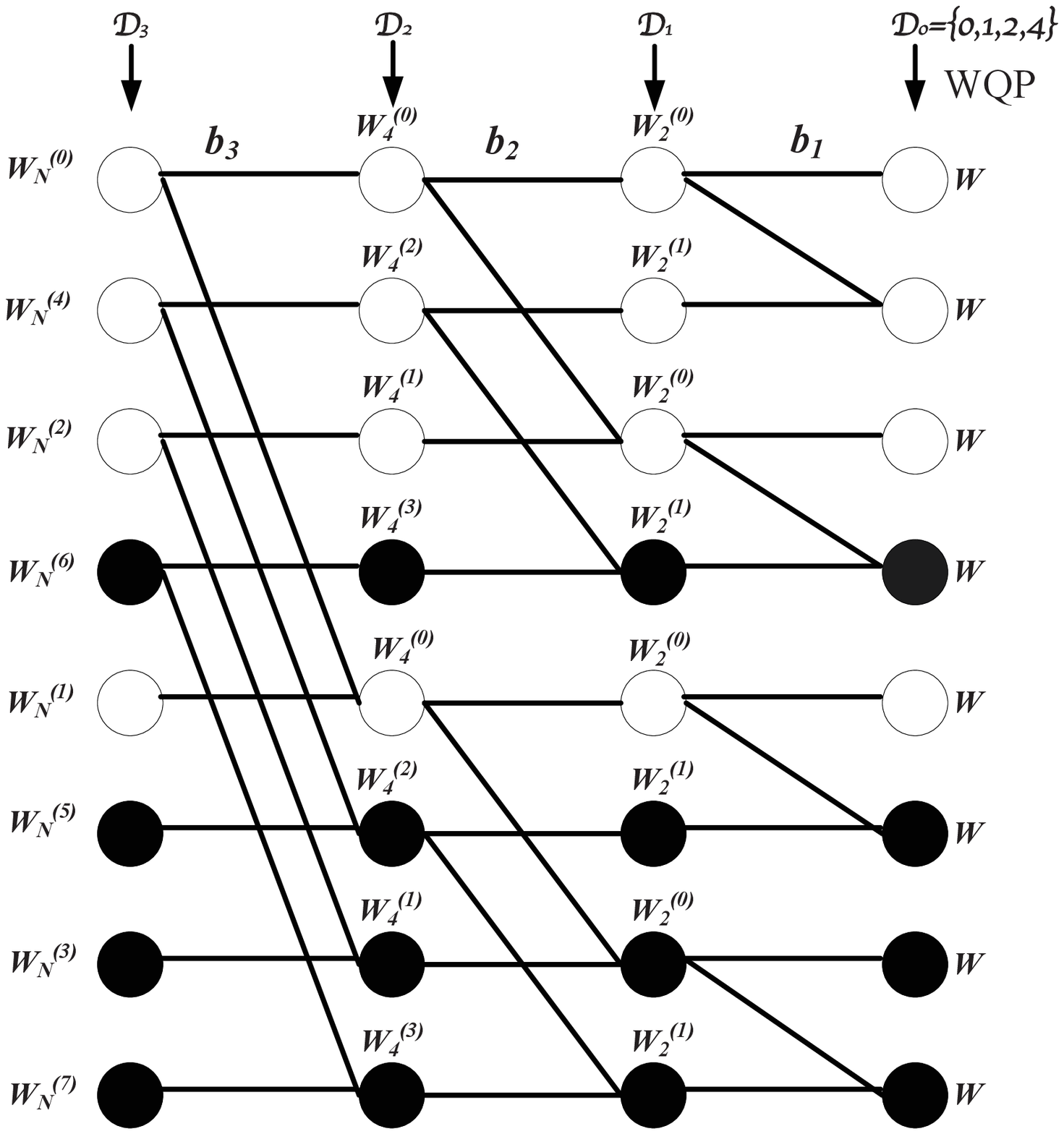}}
\label{fig_wqp}
\caption{Puncturing of QUP and WQP for $N=8$ and $Q=4$.}
\label{fig_qup_wqp}
\end{figure}

{\color{black}{
The following toy example shows the specific  process for QUP and WQP and the final punctured
bit channels of them. Let $N=8$, and $Q=4$ coded symbols are to be punctured. The QUP selects $\mathcal{P}'=\{0,1,2,3\}$
as the initial set. Then the bit reversed version of it is applied as the set for the coded
symbols, illustrated in Fig.~\ref{fig_qup_wqp}-(a), where the initial set $\mathcal{D}_0 = \mathcal{P}'$.
This procedure is channel independent. For WQP, the bit channels need to be sorted first. Assume the
underlying channel is a BEC with an erasure probability 0.5. Then the bit channel quality is sorted in the
descending order of: $[7,6,5,3,4,2,1,0]$. With $Q=4$ coded symbols to be punctured, WQP selects
$\mathcal{P}_0 = \{0,1,2,4\}$ as the initial puncturing set, shown from Fig.~\ref{fig_qup_wqp}-(b)
($\mathcal{D}_0=\mathcal{P}_0$).

Although the final reached punctured bit channels are the same for WQP and QUP in this example,
the final reached punctured bit channels could vary between these two procedures for the general cases.}}
For example, in the subsequent simulation in Section \ref{sec_numerical}, $Q=70$ coded symbols (among $N=256$) are to
be punctured. With QUP, there is actually an information bit channel $i=64$ within $[1, 70]$. This bit channel
is a punctured bit channel, resulting in an error probability of 1/2 no matter what is the operating SNR.

On the other hand, the WQP scheme searches among the worst frozen bit channels, producing punctured bit channels again in
the frozen set, as illustrated from Lemma \ref{lemma_puncture_frozen}. As long as the number of
punctured coded symbols satisfying $Q \le N-K$, there will be no punctured information bit channels from the WQP scheme.


\section{Numerical Results} \label{sec_numerical}
{\color{black}{In this section, simulations are performed for polar codes with different block lengths and code rates, also
over different channel types. Different construction procedures are also tested. Specifically, additive white Gaussian
noise (AWGN) channels and binary erasure channels (BECs) are used as the underlying channels. For
AWGN channels, the Tal-Vardy's construction procedure in \cite{vardy_it13} and the GA procedure \cite{dai_ia17}\cite{p.trifonov_c12} are employed to select
the information set. For BEC channels, the iterative calculations of Bhattacharyya parameters in \cite{arikan_it09}
are performed to select the information set. For comparison, the PW construction with $\beta = 2^{1/4}$ in \cite{zhou_vtc18}
is also used in the construction.}}
The information set is fixed once selected.

\begin{figure}
{\par\centering
\resizebox*{3.0in}{!}{\includegraphics{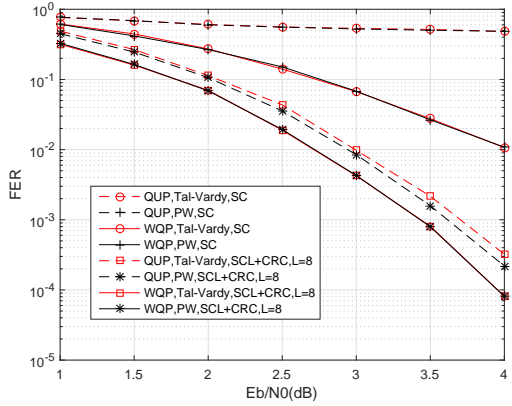}} \par}
\caption{The frame error rate (FER) of polar codes in AWGN channels. The original code length is $N=256$. After puncturing,
the code length is $M=186$ with the code rate $R=1/2$. The list size of the SCL is 8 and 8 CRC bits are added.
}
\label{FER}
\end{figure}

\begin{figure}
{\par\centering
\resizebox*{3.0in}{!}{\includegraphics{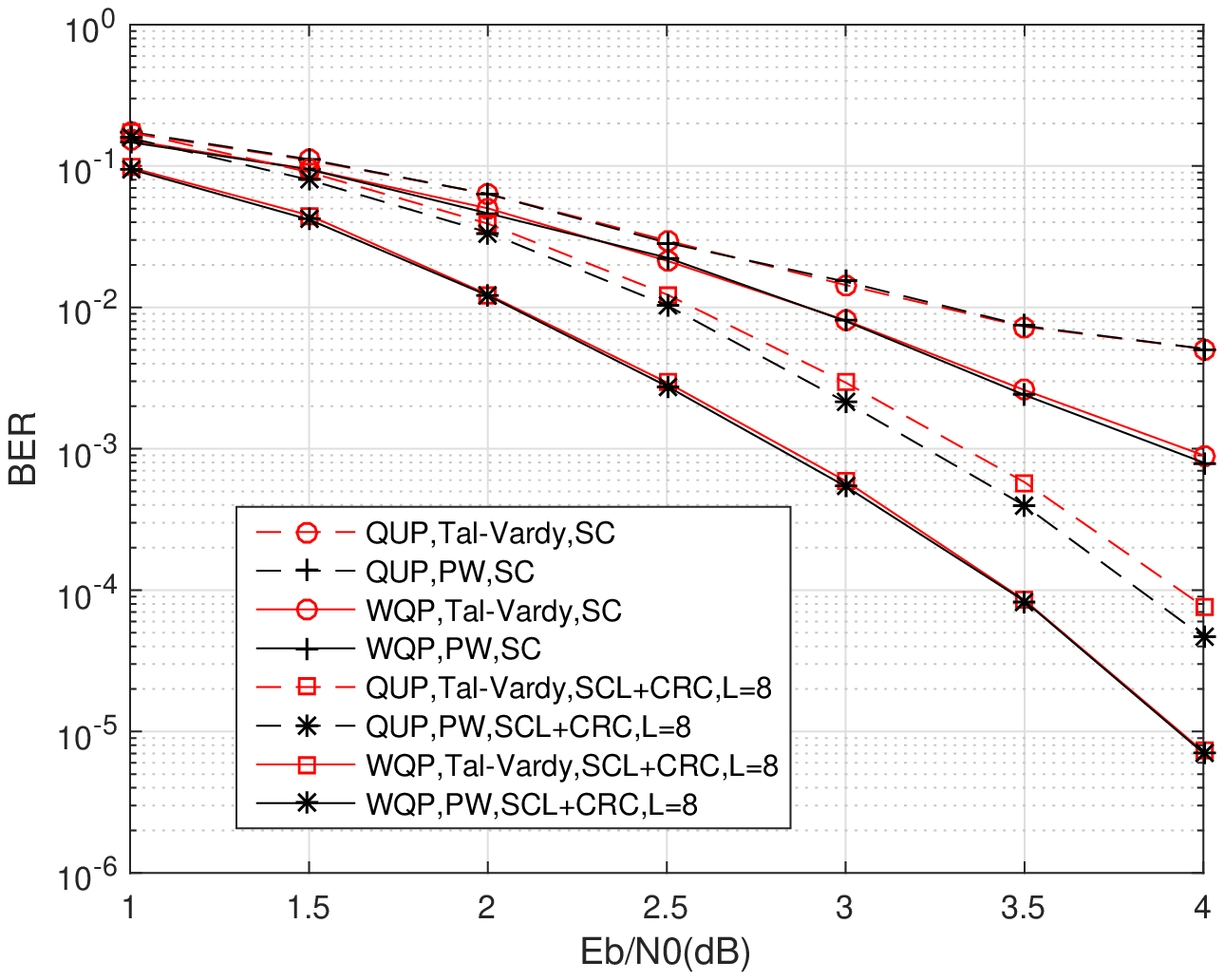}} \par}
\caption{The bit error rate (BER) of polar codes in AWGN channels. The original code length is $N=256$. After puncturing,
the code length is $M=186$ with the code rate $R=1/2$. The list size of the SCL is 8 and 8 CRC bits are added.
}
\label{ber_r_1_2}
\end{figure}

In Fig.~\ref{FER}, the frame error rate (FER) performance of WQP, along with
QUP puncturing, is shown. The corresponding bit error rate (BER) is shown in Fig.~\ref{ber_r_1_2}.
The block length of the polar code is  $N=256$. The number of punctured coded symbols is $Q=70$,
resulting in a final code length of $M=186$. The code rate is $R=1/2$ after puncturing. The SC decoding and the SCL (with CRC) decoding \cite{ vardy_it15} are both employed
in the simulations. The list size of the SCL decoding is eight, and eight CRC bits (with
the generator polynomial 0x9B) are added in addition to the original information bits.

For the SC decoding, it can be seen that WQP outperforms QUP scheme in this example, as
there is a punctured information bit channel $i=64$ for QUP. It can also be observed that,
the PW construction and the Tal-Vardy construction achieve  the same FER and BER performance with the SC decoding.

For the SCL decoding with CRC checks, the FER of the QUP scheme improves dramatically: no
error floor is observed. This is due to the eight lists and the CRC checks, which enables the decoder
to pick a correct path. The FER and BER performance of the WQP scheme also improve when it comes to
the SCL decoding with CRC checks. Again, the WQP scheme outperforms the QUP scheme with the SCL decoding.

\begin{figure}
{\par\centering
\resizebox*{3.0in}{!}{\includegraphics{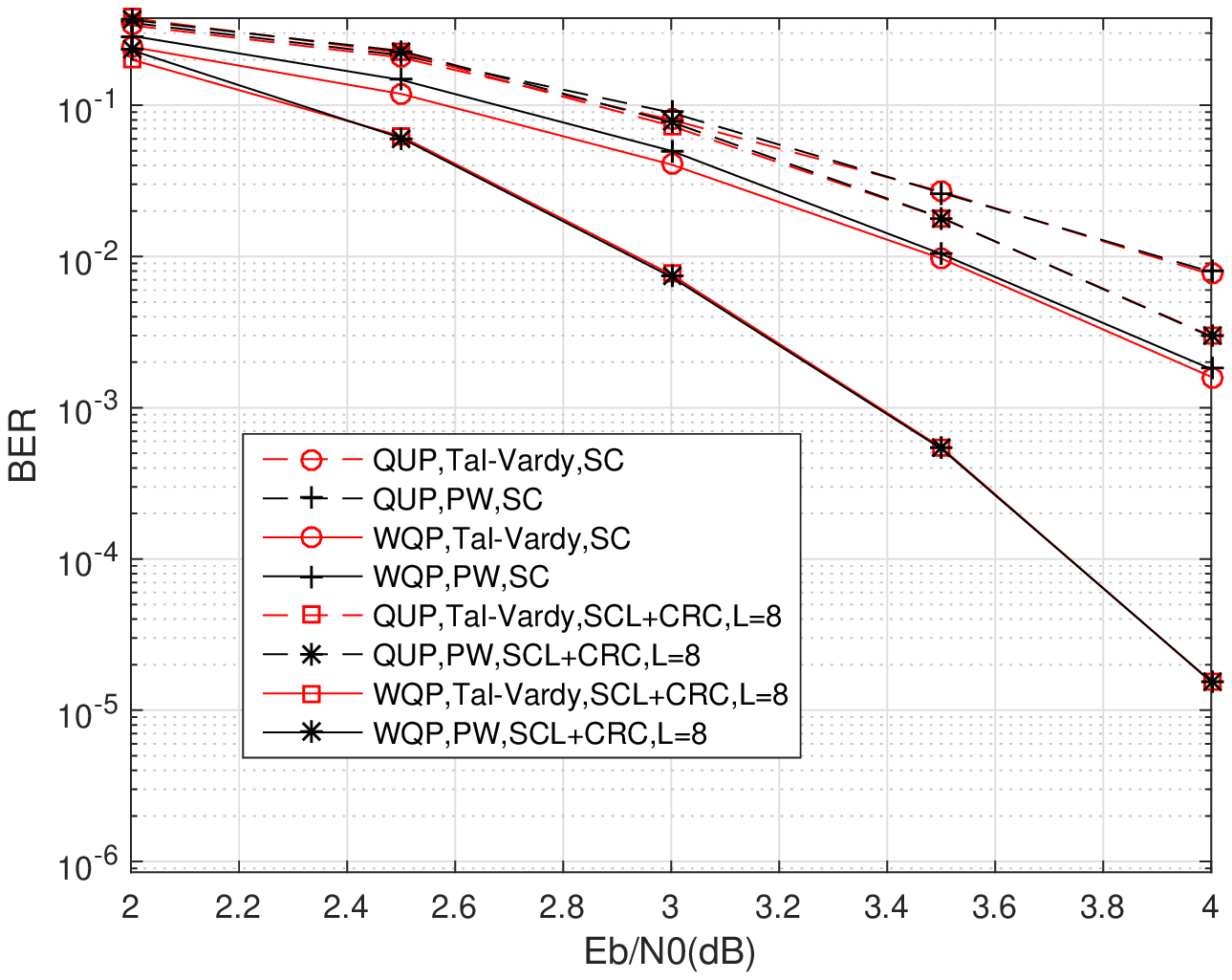}} \par}
\caption{The BER of polar codes in AWGN channels. The original code length is $N=1024$. After puncturing,
the code length is $M=904$ with the code rate $R=3/4$. The list size of the SCL is 8 and 16 CRC bits are added.
}
\label{ber_r_3_4}
\end{figure}

\begin{figure}
{\par\centering
\resizebox*{3.0in}{!}{\includegraphics{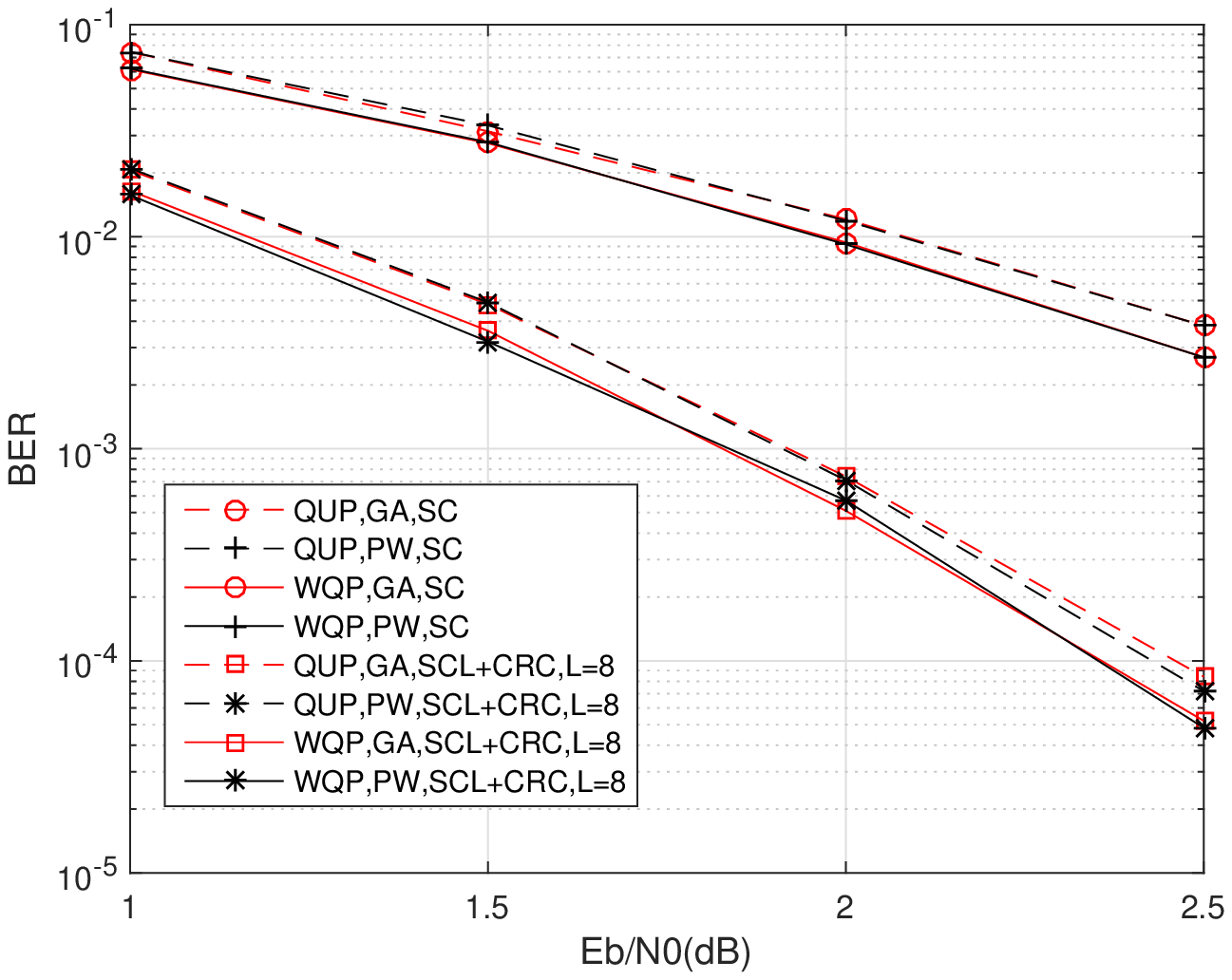}} \par}
\caption{The BER of polar codes in AWGN channels. The original code length is $N=512$. After puncturing,
the code length is $M=400$ with the code rate $R=1/4$. The list size of the SCL is 8 and 8 CRC bits are added.
}
\label{ber_r_1_4}
\end{figure}

\begin{figure}
{\par\centering
\resizebox*{3.0in}{!}{\includegraphics{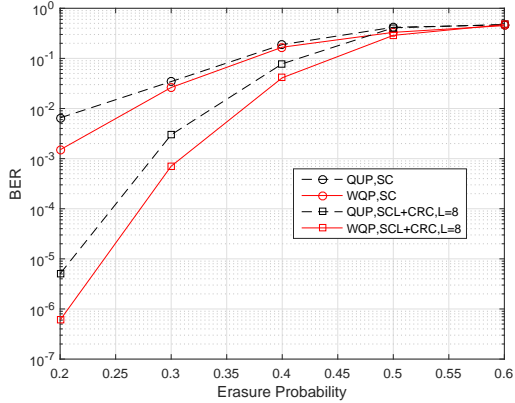}} \par}
\caption{The BER of polar codes in BEC channels. The original code length is $N=256$. After puncturing,
the code length is $M=186$ with the code rate $R=1/2$. The list size of the SCL is 8 and 8 CRC bits are added.
}
\label{ber_r_1_2_bec}
\end{figure}

{\color{black}{
The same phenomenon is observed when we increase the code length to $N=1024$, as shown in Fig.~\ref{ber_r_3_4}.
In this case, $Q=120$ coded bits are punctured. The final code rate is $R=3/4$ after puncturing.
There are 16 CRC check bits in this case, with a generator polynomial 0x8005. The
advantage of WQP is more pronounced with the increase of the code rate, since there are more
information bits to experience punctured channels for QUP. This
can be verified by the bigger gap between the performance of WQP and QUP, whether employing the SC or the
SCL decoding. When the code rate is small, the performance difference of WQP and QUP is expected
to be negligible, seen from Fig.~\ref{ber_r_1_4}. In the case of Fig.~\ref{ber_r_1_4}, the code length is
512, and $Q=112$ coded bits are punctured. The final code rate is $R=1/4$. It can be seen
that the performance between WQP and QUP is close. Note that the construction of polar codes in this case
is the GA construction, replacing the Tal-Vardy construction in Fig.~\ref{FER} to
Fig.~\ref{ber_r_3_4}.

The WQP scheme is also studied in BEC channels, reported in Fig.~\ref{ber_r_1_2_bec}. Here the parameters
of the polar code are the same as in Fig.~\ref{FER}. The performance of WQP outperforms  QUP, as in the AWGN
channels.}}

\section{Conclusion}
This paper focuses on the puncturing design of  polar codes when the information set is fixed.
The WQP algorithm is shown to be the optimal puncturing pattern, which is proven to minimize the
overall bit channel quality loss at the punctured positions. Simulation results confirm that
WQP outperforms the existing puncturing schemes when the information set is fixed.
\bibliography{ref_polar}
\bibliographystyle{IEEEtran}

\end{document}